\documentclass
{aa} 
\usepackage{graphicx}

\begin{document}

%%\title{Discs orientation in PMS binary stars}
\title{Disc orientations in pre-main-sequence multiple systems} \subtitle{A study in
southern star formation regions\thanks{Based on observations
collected at the European Southern
Observatory, Paranal, Chile (ESO Program 63.I-0358)}}
\author{Jean-Louis Monin \inst{1,2} \and Fran\c{c}ois M\'enard
\inst{1} \and Nicolas Peretto \inst{1,3}}

\institute{Laboratoire d'Astrophysique de Grenoble, UMR UJF-CNRS 5571, 
Observatoire de Grenoble, Universit\'e Joseph Fourier, BP 53, F-38041 
Grenoble cedex 9, France 
\and 
Institut Universitaire de France 
\and 
CEA/DSM/DAPNIA, Service d'Astrophysique, CEA Saclay, 
91191 Gif-sur-Yvette Cedex, France} 
\date{Received date / Accepted date} 
\offprints{J.-L. Monin,\email{Jean-Louis.Monin@obs.ujf-grenoble.fr}}
\date{Received {\today} /Accepted }

\titlerunning{Disc orientations in PMS multiple systems}
\authorrunning{Monin, M\'enard \& Peretto}

\abstract{Classical T Tauri stars are encircled by accretion discs
most of the time 
unresolved by conventional imaging observation.  However,
numerical simulations show that unresolved aperture linear polarimetry
can be used to extract information about the geometry of the immediate
circumstellar medium that scatter the starlight. Monin, M\'enard \&
Duch\^ene (1998) previously suggested that polarimetry can be
used to trace the relative orientation of discs in young binary
systems in order to shed light on the stellar and planet formation
process.  In this paper, we report on new VLT/FORS1 optical linear
polarisation measurements of 23 southern binaries spanning a range of
separation from $0.8''$ to $10''$.  In each field, the polarisation of
the central binary is extracted, as well as the polarisation of nearby
stars in order to estimate the local interstellar polarisation.  We
find that, in general, the linear polarisation vectors of individual
components in binary systems tend to be parallel to each other. The
amplitude of their polarisations are also correlated.  These findings
are in agreement with our previous work and extend the trend to
smaller separations. They are also similar to other studies, e.g.,
Donar et al.  1999; Jensen et al.  2000, 2004; Wolf et al. 2001. However,
we also find a few systems showing large differences in polarisation
level, possibly indicating different inclinations to the line-of-sight
for their discs.  
}

\maketitle

%-INTRODUCTION-----------------------------------------------------

\section{Introduction}
\label{sec:intro}
Observational studies of low-mass stellar formation show that a large
fraction of T~Tauri stars (TTS) form in binary or multiple systems
(e.g., Ghez, et al.  1993; Leinert et al., 1993; Simon et al.  1995;
Ghez et al.  1997; Padgett et al.  1997).  Theoretical studies have
shown that fragmentation appears as the most likely binary formation
mechanism to meet the observational constraints (e.g., Bate,
2000). Fragmentation mechanisms include fragmentation of a molecular
cloud core (e.g., Pringle 1989) and growth of an instability in the
outer parts of a massive circumstellar disc (e.g., Bonnell 1994).  In
the first case, neglecting long term tidal interactions, fragmentation
could yield non-coplanar systems provided that the initial cloud is
elongated and the rotation axis oriented arbitrarily with respect to
the cloud axis (Bonnell et al.  1992). In the second case, the discs
around both binary components will always be co-planar, thus the
stellar spin axes aligned. The outcome of the fragmentation process
depends on the initial conditions in the cloud and so do the final
orientations of the rotation axes of the discs in binary systems. Most
published theoretical fragmentation calculations have produced aligned
discs, but with adequate initial conditions, misaligned systems are
also a possible outcome (Bate et al. 2000).

Measuring this simple geometrical parameter of young binary systems,
the relative orientation of the discs, is important to disentangle
between various formation models.  For example, it can provide very
useful constraints on the initial distribution of angular momentum in
the parent pre-stellar cores.
  
Unfortunately, circumstellar discs in multiple systems have been
imaged in very few cases only around TTS (HK Tau: Stapelfeldt et al.,
1998; HV Tau: Monin \& Bouvier, 2000; LkH$\alpha$~263: Chauvin et al.,
2002). In each of these systems only one disc is visible and it is
seen edge-on, a favorable orientation for detection.  However, it is
not secure to conclude on strong misalignment from these measurements
only.  Indeed, only a slight tilt of the other disc away from edge-on
can abruptly reduce its detectability as the central star becomes
visible directly.  Nonetheless, and if both components have discs in
these cases, it is possible to exclude a perfect alignment to within
$\approx$ 15\,degrees.  

On the other hand, discs are often associated with jets. In some
cases, multiple jets emerge from a common unresolved location (e.g.,
Davis et al.  1994). This may indicate the presence of multiple
sources in the center, with different disc orientations.  Apart from
these few examples, i.e., in most other systems, the individual
structure of the two components in a binary is unresolved and the
determination of the relative orientation of the discs is a difficult
challenge.

Previously, Monin et al. (1998) have proposed that individual aperture
polarisation of the PMS binary components could be used to determine
the relative orientation of CS discs projected in the plane of the
sky, even when the individual discs are not resolved.  They reviewed
the literature for polarimetric measurements on wide binaries ($>8''$)
and performed CCD imaging polarimetry on closer binaries.  Their first
results showed that discs appear to be preferentially aligned, with a
few exceptions only.  They also showed that the method is very
sensitive to contamination by interstellar polarisation (ISP) that
could mimic a common disc alignment.  Other authors have obtained
similar results in the near-IR (2.2\,$\mu m$) for the Taurus region
(Jensen et al. 2000; 2004), but their results could also be impaired
by IS polarisation.

In this paper, we present new results obtained in the optical range
with a dual beam imaging polarimeter with a large field-of-view that
allows to estimate, simultaneously, the polarisation on the objects
and on surrounding field stars, i.e., provide a simultaneous
estimation of the ISP.  We believe these new measurements are better
suited to remove the contribution of the ISP and should provide a
better view of the relative orientations of the individual components
in binary systems.  The method is recalled in
section~\ref{sec:method}, and its limitations are briefly
discussed. The observations performed with the VLT FORS1 polaro-imager
(Appenzeller et al. 1998), and the data reduction process are
presented in section~\ref{sec:obs-vlt}.  The results and a discussion
are provided in section~\ref{sec:results} \and
section~\ref{sec:discussion}.

%% End INTRO ------------------------------------

\section{Determining disc orientation from linear polarimetry}
\label{sec:method}
\subsection{The method}
The method was presented in details by Monin et al. (1998): models of
disc and bipolar reflection nebulae by Bastien and M\'enard (1990)
show that the position angle of the integrated linear
polarisation of the scattered starlight is parallel to the equatorial
plane of the disc, provided that its inclination is sufficiently large
to mask the direct light from the star.
% Whitney \& Hartmann (1992) have 
% produced polarisation maps showing that when discs are surrounded by 
% envelope, the integrated polarisation is also parallel to the disc 
% plane. 
The method is thus likely to give good results when circumstellar
discs are simultaneously present around both stars in a binary, i.e.
when both are Classical TTS (CTTS) and we have restricted our study to
binaries where at least the primary is a known CTTS and/or an emission
line star. This is justified because in most T Tauri pairs, when one
of the components has an active disc, so has the other (see, e.g.,
Prato \& Monin 2000, and references therein), with mixed pairs
(CTTS+WTTS) being rare.

\subsection{The contamination by interstellar polarisation}
Interstellar polarisation is the main limitation to estimate the
intrinsic polarisation of young objects because they are found in
molecular clouds. As such, they are subject to superimposed
polarisation from the cloud they are embedded into as well as from the
interstellar medium to the observer.  When two different polarisation
directions are measured for the components of a binary system, it is
fairly secure to say that they are intrinsically different. However,
when they are similar, there is a chance that this identity is due to
a common interstellar polarisation.  Previous studies of disc
alignment in binaries have tried to estimate the local ISM
polarisation pattern from measurements found in the literature (Monin
et al.  1998, Jensen et al.  2000; 2004).  However, these estimations
rely on few measurements made at different wavelengths, at different
epochs, and sometimes quite far away from the binary under scrutiny.
%Moreover, these results only use the 
%general trend of the MIS
%polarisation orientation to deduce that the 
%binary polarisation is
%or is not oriented in the same direction and 
%can or cannot be
%trusted.  

In this paper, we have used a polaro-imager with a large field-of-view
that can simultaneously measure the polarisation of the binary and of
numerous nearby field stars. It is thus possible to estimate the
local interstellar polarisation pattern around each source studied
in this paper, under the assumption that the majority of these nearby
probes are intrinsically unpolarised.

\subsection{Measuring projected angles only}
It should be noted that the method described in this paper can only
determine the orientation of the disc projected on the plane of the
sky, i.e, its position angle. The inclination angle of the source has
no effect on the polarisation position angle, only on the polarisation
amplitude. A full determination of the relative 3D orientations of
discs in a binary system would require complementary observations of
rotational periods and $V\sin i$, or direct images, which it outside
the scope of this paper. However, Wolf et al. (2001) have shown that
this problem can be addressed statistically. They showed that the
probability distribution function of position angle differences will
peak toward zero if discs have a tendency to be aligned. It remains
possible however, for a given binary, to assess that its discs are not
aligned when the PA difference is large.

\section{Observation and data reduction}
\label{sec:obs-vlt}
\subsection{Source selection}

The sources we studied are taken from the list of Reipurth \&
Zinnecker (1993, RZ93). The same source names are used.  The angular
separation of the binaries ranges from $0.8''$ to $10.6''$,
corresponding to linear separations from 70 to 1900~AU, assuming the
distance values given in RZ93, and Geoffray \& Monin (2001) for Hen~3-600. 
The binaries were chosen in various
southern star formation regions (SFR) with the condition that at least
the primary is a known CTTS or emission line star.  They are listed in
Table~\ref{tab:src-params}, sorted by SFR of increasing right
ascension, and, within a given SFR, by increasing separation. We
keep this classification order throughout the rest of the paper.

% Then most of the time, the secondary will be of the same TTS
% type as the primary, and polarimetry should give access to the disc
% orientation for both components aswell.

\begin{table}[htb]
\caption[]{\label{tab:src-params} Source parameters, listed by SFR of
increasing RA, and increasing angular separation within a given SFR.}
\begin{flushleft}
\begin{tabular}{llllll}
\hline\hline
Source & HBC &  SFR&  Sep ($''$) & Sep (AU)  
%& $N_{\rm cont}$ 
\\
ESO\,H$\alpha\,29$ &   & Gum  &  4.2 &  1900 \\
Hen 3-600 &     &TW~Hya &  1.5  & 70    \\
Sz 30 &   & Cha I  &1.2 & 170    \\
Sz 2 & 564 & Cha I & 2.2 & 310  \\
Glass-I &     & Cha I & 2.4 & 340   \\
Sz 15 &   &  Cha I &10.6 & 1500   \\
Sz 48 &  & Cha II & 1.31 &  260    \\
Sz 62 &   & Cha II&  1.7 & 330  \\
Sz 60 &   & Cha II  &3.4  & 670 \\
HO Lup & 612 & Lup & 1.5 & 220    \\
Sz 116 & 625 & Lup  & 1.6 & 240    \\
SZ 81 & 604 &  Lup  & 1.9 &  285   \\
SZ 65 & 597 &Lup  & 6.4 &  960    \\
WSB 20 &   & Oph &0.8  & 130     \\
WSB 18 &   & Oph & 1.1 & 170    \\
WSB 26 &   & Oph & 1.2 & 130   \\
WSB 19 &   & Oph& 1.5  & 24 \\
WSB 35 &   & Oph & 2.3 & 360    \\
WSB 4 &   &Oph  & 2.8  &  450   \\
WSB 42 &  & Oph &   5.1&  820    \\
WSB 28 &   & Oph & 5.1 & 1400    \\
HBC 679 & 679 &CrA & 4.5 & 580   \\
AS 353 &  292  &L673  & 5.7  & 1700 \\
%

%%AS 205 A & 1.29 & 1.45& 0.20& 88.9 &2.9 \\ B & & 5.53& 1.06& 51.3
%%&4.6 \\ VV CrA A & 1.93 & 4.18& 0.25& 108 &1.7 \\ B & & 6.78& 1.73&
%%104.2 &7.3 \\
\hline
\end{tabular}
\end{flushleft}
\end{table}
The measure the interstellar polarisation near the targets of our
study, we use background stars.  We have estimated the number of {\em
foreground} stars that can be expected in a FORS1 field-of-view (see
\S~\ref{sec:obs}) toward each target is very small.  We have used the 
galactic model from Bahcall \& Soneira (1984) to compute how many
stars should be present in the field in front of every source, given
its galactic coordinates and distance.  The result shows that in all
cases but two, the foreground contamination is less than one star per
field. Therefore, we use the stars present in each field to estimate
the interstellar polarisation. The foreground-contaminated sources are
ESO\,H$\alpha$\,29 (20 possible foreground objects) and AS\,353 (7
objects).

Note that the polarisation of background stars is distance dependent
(e.g., Serkowski et al., 1975). We suggest a method to test the
reliability of our estimate in section \ref{subsec:binvsims}.
However, the regular pattern as well as the uniform degree of the
polarisation observed around many of our sources suggest that the
interstellar polarisation originates from a slab of dust (presumably
that of the molecular cloud) rather than the diffuse interstellar
medium (see section~\ref{sec:results} and Figure~\ref{fig:cartepol1}
and \ref{fig:cartepol2}).

%In order to confirm the status of CTTS for most of the members of our
%list, we have recovered their 2MASS JHK data. 
%Figure~\ref{fig:2mass-data} shows the J-H/H-K diagram for our sources. 
%To a few exceptions, all sources are consistent with being on the CTTS
%locus under more or less supplementary visual absorption.
%
%   \begin{figure}[htb]
%    \begin{center}
%\includegraphics[width=8 truecm]{jh-hk-2mass.eps}
%\caption[]{\label{fig:2mass-data}J-H/H-K diagram of all the
%(unresolved) sources of our sample, obtained from 2MASS. The
%diamond-dotted sequences are for MS stars and giants, the dashed line
%is the locus of CTTS from Meyer et al.  (1997), and the thick vector
%draws an $A_{V}=10$ reddening.  Most of the sources show NIR excess
%indicative of discs}
%\end{center}
%\end{figure}
 %
\subsection{Observations}
\label{sec:obs}

Observations were made in the I-band during 5 nights on 2000 May
24-29.  The weather conditions were good and the seeing was measured
between 0.5 and $1.7''$ with a median value of $0.7''$ over the 5
nights.  A good seeing is important because it sets the effective
separation down to which binaries can be resolved. Integration times
between 0.5~s and 2~min were used depending on the brightness of and
the contrast needed in, each binary system.

The FORS1/IPOL instrument is equipped with a Wollaston prism that
splits the incident beam in two different directions with orthogonal
polarisation states, the so-called ordinary ($o$) and extraordinary
($e$) beams.  A stepped half-wave plate retarder is placed at the
entrance of the incident beam and can be rotated, in this case by
multiples of 22.5$^{o}$ so that 16 positions are needed for a complete
rotation. The separation of the two $o$ and $e$ beams on the CCD is
performed by the Wollaston prism and a 9-slit focal mask. Each slit is
$\sim20''$ wide.  For each position of the rotating retarder plate, an
image is recorded. The images are then combined to yield the Stokes
parameters I, Q and U.
%We
%could have measured a complete square field of view using two
%measurement sets at two positions on the sky separated by an
%inter-slit offset, but our main goal was to measure the central binary
%and a sufficient number of field stars, and we have just kept
%shaded-like images.  

The total field-of-view of FORS1/IPOL is $6.8'\times 6.8'$ in the
Standard Resolution (SR) mode with a focal scale of $0.2''$/pixel.  To
obtain the polarisation, the normalized flux difference between the
ordinary and extraordinary images, either from aperture photometry on
point sources or pixel by pixel on extended objects, was calculated
and a Fourier series computed to derive the Stokes parameter Q and U\footnote{See the FORS user manual at http://www.eso.org, and also Patat
\& Romaniello (2005). The polarisation level, $P$, is obtained by calculating $P=\sqrt{Q^2+U^2}$ and the position angle, $\Theta$,  by calculating
$\Theta = 1/2\arctan(U/Q)$.}
%In every case, an unpolarised test
%exposure was first acquired to adjust the individual exposure time in
%order to maximize the signal/noise ratio while allowing the peak
%signal to vary within the detector dynamics due to seeing
%fluctuations.  
In aperture photometry, with an aperture 3 FWHM in size and using all
16 rotations of the waveplate (yielding 8 independent estimations of
$Q$ and $U$), the error on the polarisation is estimated at 0.1~\% or
better from photon noise only.
 
% Most of the time, we obtained measurements in I and V while some of
% the sources have only been measured in I. The main scope of this paper
% is to study disc alignments, so that when both bands are available, if
% no real difference appears between the I and V measurements, we give a
% global result.  When necessary, we discuss the difference.

%%\section{Data Reduction}
\subsection{Data Reduction Pipeline}
A dedicated data reduction pipeline was written using NOAO/IRAF.  The
images are first bias and bad pixel corrected, and then flat-fielded.
In the next step the images go through a polarisation extraction
routine.  Two options are then available: the polarisation information
can be estimated on a pixel per pixel basis, a useful possibility to
map extended structures like reflection nebulosities, at the cost of a
loss of accuracy on point sources when the image quality (FWHM)
changes during acquisition of a full data set (i.e. between different
positions of the half-wave plate). The other option uses aperture
photometry to estimate precise polarisation measurements on point-like
objects.  In aperture photometry mode, any FWHM change can be
accounted for if a large enough aperture is used, typically 3~FWHM.
% our binary objects.  Both techniques are available in our pipeline on
% request, to estimate the polarisation of point sources and extended
% features alike.
For a few of the tightest binaries of our sample, we modeled a point
spread function from reference stars to extract the photometric
signal of the two components by PSF subtraction.
% but most of the
%time, we used the aperture photometry method to extract the
%polarisation in our binary objects.
% Then a
% Fourier series is computed to extract the Stokes parameter U \& Q,
% hence P and $\theta$ from the data.

The errors were estimated using 2 independent methods: first, from the
photon noise on the $e-$ and $o-$beams separately, and then
propagating the errors in the calculations of Q, U, P and $\Theta$;
second, by measuring the standard deviation on the 4, 8 or 16 images
from the half-wave plate rotation.  Both estimations give consistent
results except in a few pathological cases like, e.g., severe hit by
a cosmic ray, sources too close to dead zones between orthogonal
polarisation strips, etc. 

The estimated error is less than $\sigma$(P)=0.1\% (absolute value)
when the binary components are well separated ($\ge$1.3
arcsec). However, in Table~\ref{tab:pol-results} and in subsequent
computations, we conservatively use the worse value of the error
estimated from the two methods. The resulting signal to noise ratio on
the measured polarisation is usually high ($P/{\sigma} > 10$) and we
present our results without correction for low signal to noise bias
(e.g., Wardle \& Kronberg, 1974).

\subsection{Instrumental polarisation at the center of the FORS1 field}
Of crucial importance is the determination of the instrumental
polarisation, $P_{\rm inst}$.  We have carefully estimated it by
measuring nearby (i.e., high proper motion) unpolarised targets.  We
have observed GJ~781.1 and GJ~2147, two high proper motion stars.
Because the immediate solar neighborhood is remarkably devoid of dust,
the interstellar polarisation of nearby stars can be considered null.
The average of 4 measurements on both GJ objects gives $P_{\rm
inst}=0.02\%\pm 0.03\%$. As a further check of very low instrumental
polarisation at the center of the field, many binaries in our sample
have low linear polarisation, known from previous publications. Our
measurements with FORS1 are very close to previously published data.
We therefore believe that FORS1/IPOL instrumental polarisation is very
low on-axis, well below 0.1\% at the center of the field, and we did not
attempt to remove it from the measurements. We address the case of the
spatial dependence of the instrumental polarisation in
section~\ref{sec:poldep}.

\section{Results}
\label{sec:results}

\subsection{Polarisation data}

\begin{figure*}[h]
    \begin{center}
\includegraphics[width=16 truecm]{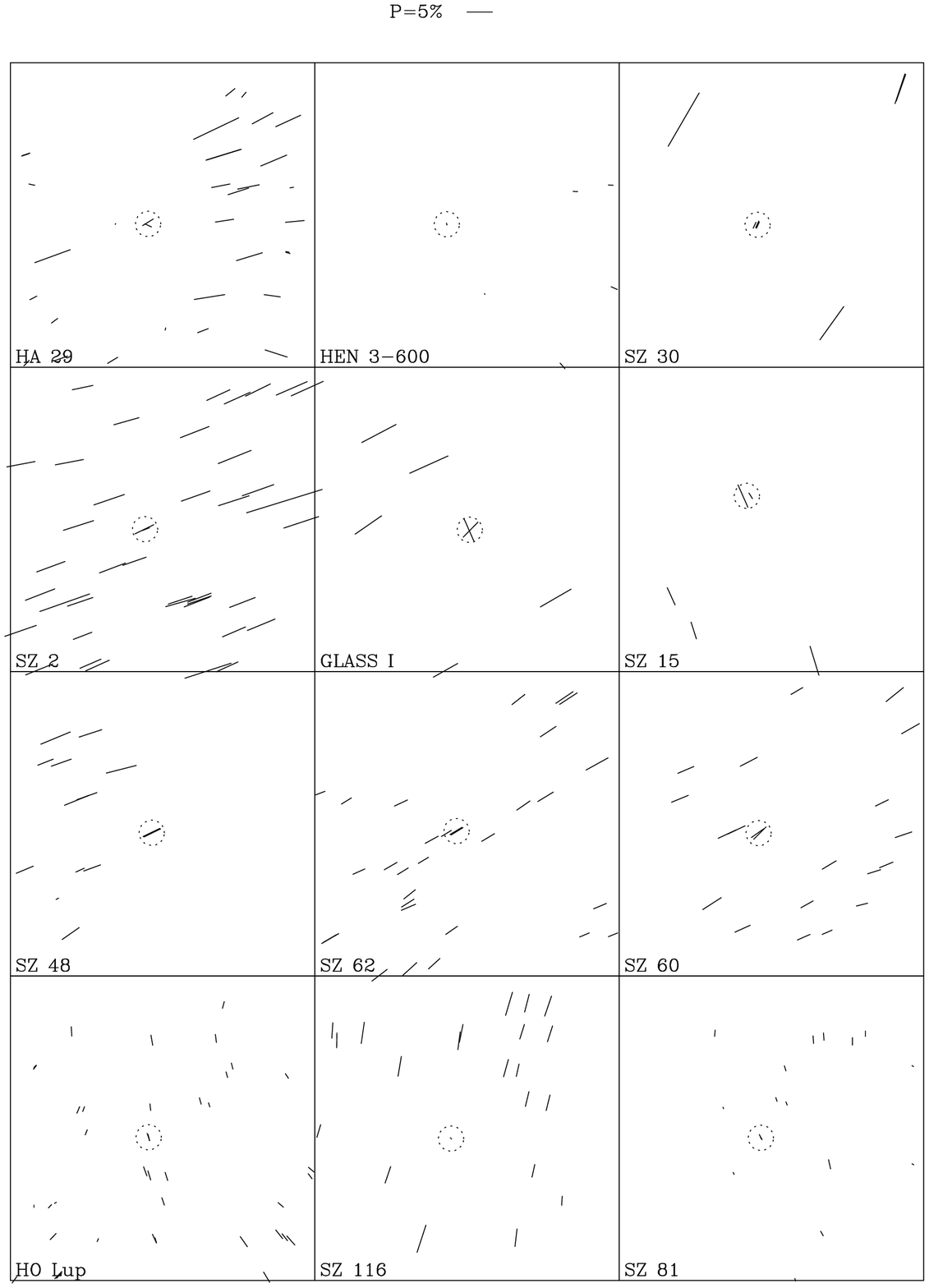}
\caption[]{\label{fig:cartepol1} I band polarisation maps for the
first 12 sources in our list.}
\end{center}
\end{figure*}

\begin{figure*}[h]
    \begin{center}
\includegraphics [width=16 truecm]{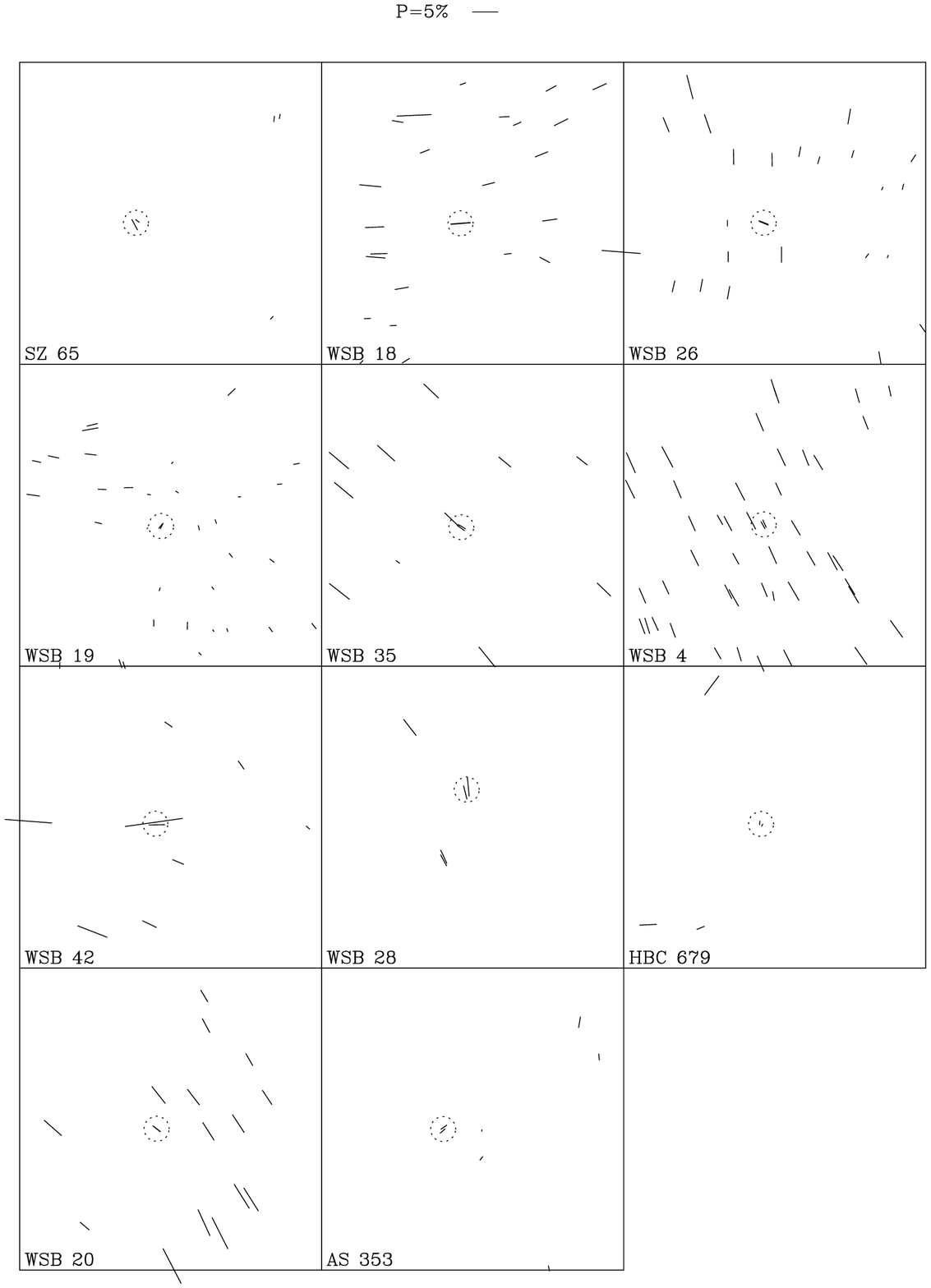}
\caption[]{\label{fig:cartepol2}  I band polarisation map of the last 
11 sources in our list.}
\end{center}
\end{figure*}

For each field observed, the polarisation level, $P$, and the position
angle, $\Theta$, is measured for every star for which
S/N$\ge 1000$ ($\sigma_P=0.1\%$).   Figures~\ref{fig:cartepol1} and
\ref{fig:cartepol2} show the polarisation maps obtained around the 23
binaries listed in Table~\ref{tab:src-params}.  Across most of the
fields, the polarisation presents a smooth pattern, both in $P$ and
$\Theta$.  However, in some of the fields the polarisation appears
more chaotic, in amplitude and/or in position angle; this is the case
for instance around WSB\,19.


Table~\ref{tab:pol-results} lists the values computed for $P$ and
$\Theta$ on the central binary components and on the surrounding
interstellar medium for all sources.  The method used to extract the
interstellar polarisation (col. 6 \& 7) is detailed in
sect.~\ref{subsec:isp-estim}.

\begin{table*}[htb]
\caption[]{Polarisation measurements of the individual components (A,B) and
estimation of the Interstellar polarisation (IS, computed in
sect.~\ref{subsec:isp-estim}, see text for details).  In every column,
the number in parenthesis gives the $1\,\sigma$ uncertainty. The three
rightmost columns list the visual extinction from the literature and
the references: (a) Prato et al.  2003; (b) Brandner \& Zinnecker
1997; (c) Geoffray \& Monin 2001. }
\label{tab:pol-results} 
\begin{flushleft}
\begin{tabular}{llllllllll}
\hline\hline
Source & P$_A$ (\%) & $\Theta_A (^o)$& P$_B$ (\%) & $\Theta_B (^o)$ &
P$_{IS}$ (\%) & $\Theta_{IS}(^o)$
 & $A_{V}(A)$ & $A_{V}(B)$ & Ref.\\
ESO H$\alpha\,29$ &  1.39 (0.06) & 65 (2) &
2.47 (0.08) & 122 (1) & 3.2 (0.1) & 105 (1) &    &   & \\
Hen 3-600  &  0.27 (0.04) & 9 (5) &
0.02 (0.05) & 171 (80) & 0.84 (0.07) & 53 (2) &  0.7   &   0.7& c \\
Sz 30  & 1.3 (0.05) & 156 (1) &
1.4 (0.05) & 155 (1) & 5.3 (0.2) &154  (1) &   0.58   &   0.19 & b  \\
Sz 2   &  1.53 (0.05) & 112 (1) &
4.10 (0.05) & 116 (0.5) & 5.4 (0.2) & 112 (1) &    &   & \\
Glass-I   &  4.15 (0.05) & 135 (1) &4.88
(0.06) & 24 (1) &7.46 (0.2) & 118 (1) &     &  &  \\
Sz 15  &  1.28 (0.04) & 32 (0.8) & 4.81
(0.15) & 24 (0.8) & 4.0 (0.15)& 22 (0.6) &    &  &  \\
Sz 48    & 3.62 (0.03) & 116 (0.3)
& 3.70 (0.04) & 116 (0.3)  & 1.29 (0.04) &120 (0.5) &    3.41   &   3.58  & b \\
Sz 62   &  2.76 (0.11) & 122 (1) & 2.72
(0.12) & 121 (1)  & 2.88 (0.09) & 121 (1) &    1.08   &   1.58  & b\\
Sz 60  &  3.47  (0.04) & 135 (1) & 2.94
(0.04) & 125 (0.5)  & 3.0 (0.1) &117 (1) &    &  & \\
HO Lup  &  1.21 (0.07) & 12 (2) &
1.51 (0.06) & 19 (1)   & 0.8 (0.03) & 7 (1) &   1.25   &   & c\\
Sz 116  &   0.04 (0.07) & 158 (46) &
0.18 (0.1) & 71 (15)  & 3.55 (0.12)& 165 (1) &    0   &   0.9 & a \\
SZ 81  &  0.4 (0.05) & 39 (4) & 1.11
(0.06) & 30 (2)  & 0.79 (0.04) & 7 (1) &  &  &   \\
 SZ 65   &  0.89 (0.06) & 53 (2) & 2.3
 (0.2) & 29 (3) & 0.84 (0.12) & 168 (4) &    &  & \\
WSB 20   &  2.84 (0.05) & 113 (1) 
& 2.85 (0.05) & 113 (1)   & 4.10 (0.04) & 124 (0.3) &  2.3   &   & c\\
WSB 18 &  3.8 (0.04)  & 95 (0.3) 
& 3.79 (0.03) & 95 (0.3)  &  2.08 (0.07)    & 99 (1)    &   4.04  &  3.41 & b\\
WSB 26   &  1.95 (0.04) & 68 (0.5) &
1.95 (0.04) & 68 (0.6)  &  1.72 (0.06) & 175 (0.5) &   &  & \\
WSB 19  &  1.09 (0.06) & 139 (2) & 1.26
(0.1) &152 (2)   &  0.42 (0.03) & 55 (2) &   1.7   &   2.7  & b \\
WSB 35 & 1.58 (0.04) & 59 (0.7) & 1.81
(0.05) & 54 (1)     & 4.34 (0.14) &48 (1) &   &  & \\
WSB 4   &   1.52 (0.04) & 31 (1) & 1.57
(0.04) & 25 (1)   & 2.75 (0.1) & 21 (1) &   0   &  0.4 & a \\
WSB 42   &   3.10 (0.04) & 91 (1) & 11.4
(0.1) & 98 (1)    & 5.2 (0.2) & 81 (1) &   &   & \\
WSB 28   &  2.67 (0.04) & 14 (0.3) &3.75
(0.3) & 5 (2)   & 2.52 (0.09) & 28 (0.5) &   5.1   &  2.5 & a\\
HBC 679   &  0.47 (0.04) & 156 (2)
& 0.73 (0.22) & 176 (9)  &  1.71 (0.06) & 106 (1)  &  4.8   &  1.6 & a \\
AS 353  &  1.40 (0.04) & 125 (1) & 1.25
(0.09) & 133 (2)   & 1.0 (0.1) & 176 (3) &   &  2.1 & a\\
%

%%AS 205 A & 1.29 & 1.45& 0.20& 88.9 &2.9 \\ B & & 5.53& 1.06& 51.3
%%&4.6 \\ VV CrA A & 1.93 & 4.18& 0.25& 108 &1.7 \\ B & & 6.78& 1.73&
%%104.2 &7.3 \\
\hline
\end{tabular}
\end{flushleft}
\end{table*}
\begin{figure}[htbp]
    \begin{center}
\includegraphics[width=8 true cm]{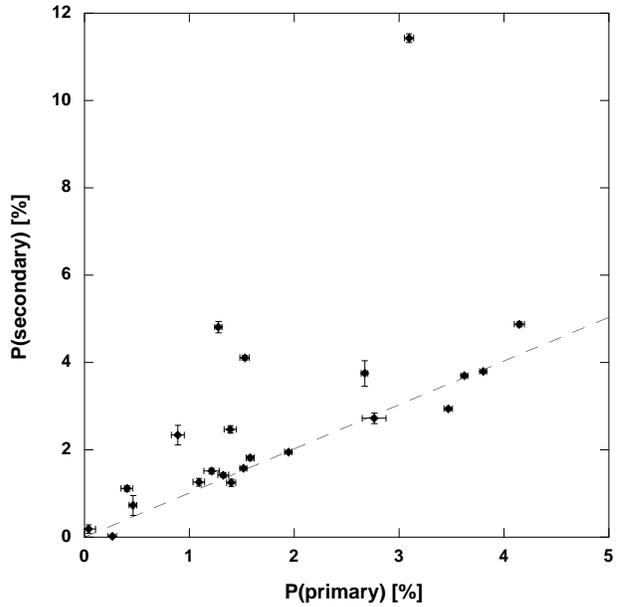}
\caption[]{\label{fig:pbvspa}  Measured polarisation level in the 
secondary versus in the primary. The dashed line traces an identical 
polarisation level in both components.}
\end{center}
\end{figure}
In Figure~\ref{fig:pbvspa}, we have plotted the polarisation level of
the secondary component against that of the primary for all the
binaries in our list except for WSB\,20 which is too tight to obtain a
reliable estimation of the individual polarisations. WSB\,20 will not
be considered in this study from now on. The plot shows that the
polarisation levels of both components in a given binary are
correlated.  This result is expected if the discs of each components
are similar in optical thickness and have the same inclination. It may
also reflect a lack of intrinsic polarisation but a common
contamination by the ISP.  At first sight, Figure~\ref{fig:pbvspa}
also suggests that the polarisation level of the secondaries' are
often larger than the primaries'. This result is hard to explain if the
measurements are dominated by the interstellar polarisation. It will
be further analyzed and discussed in section~\ref{subsec:secpol}.

\subsection{Estimation of the nearby interstellar polarisation}
\label{subsec:isp-estim}

% Depending on the ISP orientation on the central binary
% and on the surrounding field, one could expect the local ISP to cancel
% the central object polarisation, so that having no central
% polarisation while the local ISP seems strong does not imply that the
% central object is unpolarised or placed at a different position than
% the local cloud.  Therefore we will not try to guess any local
% orientation of the ISP from our polarisation maps, but rather
% concentrate on the best possible estimation of the ISP $Q$ and $U$
% parameter for later subtraction from the central source.
% To do so, we have computed the plain average value of Q and U over 
% the field of view: $1/n\Sigma Q_{i}$ (resp. U) and 
% compare it to the error-weighted average of these values: $(\Sigma 
% Q/\sigma^{2}_{Q})/(\Sigma 1/\sigma^{2}_{Q})$ (resp. U). The 
% correlation coefficient between the two estimations is very good, 
% showing that there are none but few spurious points in the measurements. 
% We therefore use the weighted average of Q and U as best estimators to 
% build the local interstellar polarisation.
In order to estimate the ISP at the center of each field, it is
assumed that none of the surrounding stars  are intrinsically
polarised. In that case, the noise-weighted averages of their $Q$ and
$U$ Stokes parameters, computed over the whole field-of-view and
excluding the central binary, can be used as an estimation of the ISP.

\begin{eqnarray}
    \label{eq:estime-q}
    \overline{Q} & = & {{\Sigma {{Q}\over{\sigma^{2}_{Q}}}}\over{\Sigma
    {{1}\over{\sigma^{2}_{Q}}}}}\\
    \label{eq:estime-u}
    \overline{U} & = & {{\Sigma {{U}\over{\sigma^{2}_{U}}}}\over{\Sigma
    {{1}\over{\sigma^{2}_{U}}}}}
\end{eqnarray}

However, estimation of the ISP based on the computation method
described in Eq~(\ref{eq:estime-q}) and (\ref{eq:estime-u}) requires
caution. For example, accurately estimated single measurements can be
spread over a large range of values (in $P$ and/or $\Theta$), possibly
from superimposed interstellar clouds at various distances along the
line of sight, and thus lead to a poorly representative estimation of
the ISP at the distance of the binary.  Therefore, further inspection
of each polarisation map is also needed to disentangle `regular' from
`irregular' ISP. Usually a quick visual inspection is sufficient. For
example, WSB\,19 shows two superimposed ISP components without any
clear trend around the central position. WSB\,19 is discarded
from the sample for now on. 

In general however, the observations show smaller fluctuations of the
ISP across the field (e.g., ESO\,H$\alpha$\,29, and HO\,Lup), and the
mean of the of the peak in the histogram of the position angles can be
used to estimate the orientation of the ISP.  In other cases, the ISP
is very well defined across the field and its evaluation is
straightforward. This is the case, e.g.,  for the fields around SZ\,2 or
SZ\,62.

%In some of our sources, the ISP PA distribution sometimes shows two
%components : a regular gaussian distribution superimposed on a flat
%spread of sources with a large range of PA. In such case, we assumed
%that the gaussian distribution traces the local ISP while the flat
%distribution is unrelated and it is discarded.
%Figure~\ref{fig:histo-ha29} illustrates this point. The PA
%distribution of the polarisation is plotted and a gaussian is fitted
%on the peak of the distribution only.
% The fit gives a gaussian peak at
% $\Theta=120^{o}$, when the weighted average of the polarisation gives
% $\Theta=105^{o}$, hence a consistent value.
% %
%\begin{figure}[htb]
%    \begin{center}
%\includegraphics[width=6 truecm,angle=270]{fit-histo.eps}
%\caption[]{\label{fig:histo-ha29} Histogram of the ISP value 
%on a chaotic field (source ESO\,H$\alpha$\,29 in 
%fig.~\ref{fig:cartepol1})}
%\end{center}
%\end{figure}

%Finally, the median of the ISP is computed for every sources.
%In the case of ESO\,H$\alpha$\,29 previously mentioned, the remaining
%individual stars are regularly spread across the field and the median
%ISP PA is very close to the average regular PA ($\pm 1^{o}$).
%A_T_ON BESOIN DE CETTE DERNIERE PHRASE?

In practice, the average value of the ISP computed in
equations~(\ref{eq:estime-q}) and (\ref{eq:estime-u}) explicitly
removes the influence of poor quality measurements (e.g., with large
errors) while the median value eliminates spurious values (possibly
measured with small errors).  When both values, median and average,
coincide the determination of the local ISP is considered reliable.
Otherwise, no attempt is made to subtract an ISP component.

\begin{figure}[htb]
   \begin{center}
\includegraphics[width=8 truecm]{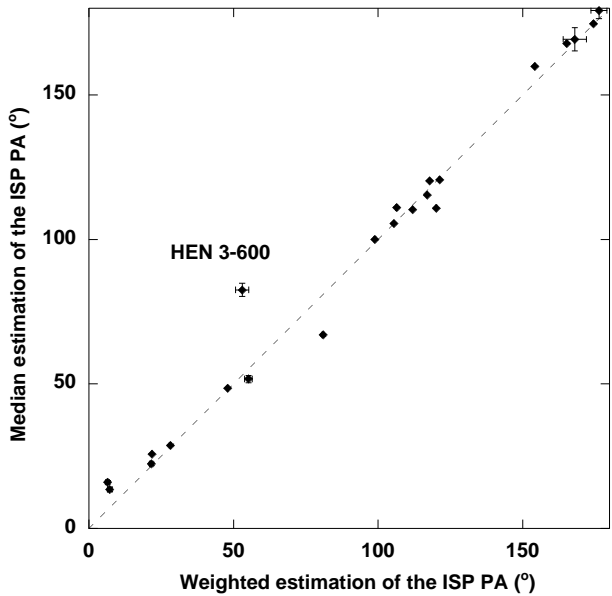}
\caption[]{\label{fig:correl-d-p} Median versus weighted Position 
Angle estimation of the interstellar polarisation for all sources in Table~\ref{tab:src-params}
except WSB\,20.}
\end{center}
\end{figure}

\begin{figure}[htb]
   \begin{center}
\includegraphics[width=8 truecm]{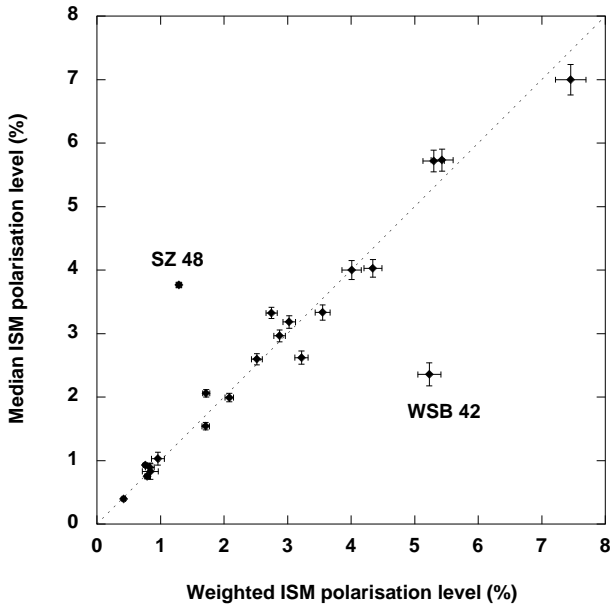}
\caption[]{\label{fig:correlP-d-p} Median versus weighted percentage
level estimation of the interstellar polarisation for all sources in Table~\ref{tab:src-params}
except WSB\,20.}
\end{center}
\end{figure}

In Figure~\ref{fig:correl-d-p} and \ref{fig:correlP-d-p}, we have
plotted the weighted estimate of the ISP PA and percentage level
vs the median value, showing that both estimates give the same
result, except for 3 sources (SZ\,48, WSB\,42 and Hen\,3-600). 

As the weighted average is best evaluated from a signal-to-noise 
point of view, we keep it for further ISP subtraction.  We do not
attempt to subtract an interstellar component from the 3 discordant
sources, 18 sources remain in the sample.

\subsection{Binary vs. interstellar polarisation}
\label{subsec:binvsims}

\begin{figure}[htbp]
    \begin{center}
\includegraphics[width=8 true cm]{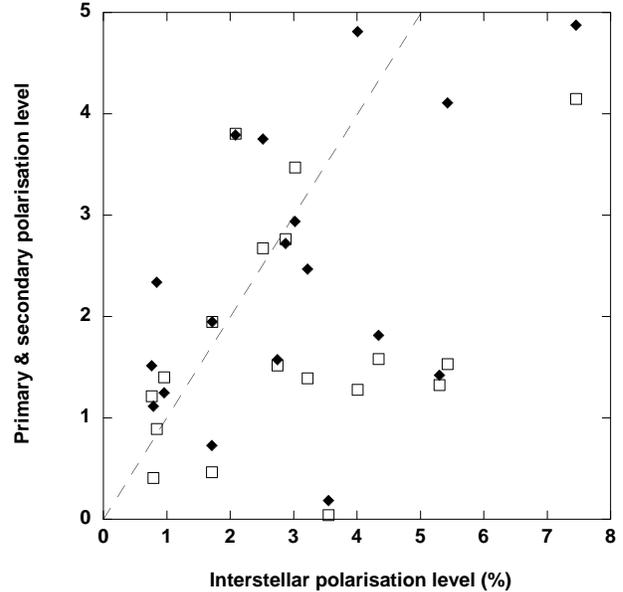}
\caption[]{\label{fig:pabvspmis}  Polarisation level of the binary 
components (primary: open squares; secondary: full diamonds) 
vs interstellar polarisation for the 18 sources where the ISP is best 
evaluated. The dashed line traces the locus of fully influenced sources.}
\end{center}
\end{figure}

\begin{figure}[htbp]
    \begin{center}
\includegraphics[width=8 true cm]{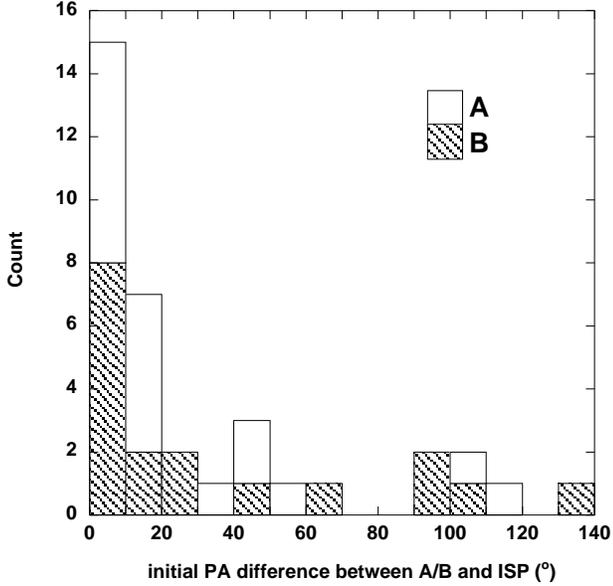}
\caption[]{\label{fig:paabvspamis} Distribution 
of the polarisation angle difference on
18 binary components vs interstellar polarisation.  Primaries are 
open squares, secondaries are shaded.}
\end{center}
\end{figure}

Before subtracting the local ISP, we consider in this section the
results from the raw polarisation
measurements. Figure~\ref{fig:pabvspmis} shows that there is no strong
correlation between the interstellar polarisation and the 
polarisation levels of the individual components.
Similarly, Figure~\ref{fig:paabvspamis} shows the histograms of the
position angle difference between the components and the interstellar
polarisation for the primary and the secondary in the 18 sources with
individual measurements and a reliable estimate of the ISP. Two thirds
of the sources show both polarisations parallel to the ISP, to within
30 degrees (20 degrees for the primaries).  Yet, one third of the
sources are not aligned with the ISP. It indicates that at a fraction
of their observed polarisation is intrinsic, different from the
estimated ISP. On these sources, the results can be used to trace the
actual orientation of the discs.

%We use figure~\ref{fig:paabvspamis} to split our 18 sources in two
%groups.  The first group contains 6 binaries where both the primary
%and the secondary have their polarisation oriented at more than 30 deg
%from the ISP and are therefore {\em a priori} not influenced by the
%interstellar polarisation.  The second group contains the 15 residual
%binaries where at least one of the components shows a polarisation
%oriented within less than 20 deg from the ISP; among these 15 sources,
%11 have both components oriented close to the ISP. These latter
%sources are thus suspected to appear aligned via the influence of the
%ISP, so that we can not extract significant results from
%the direct comparison between the two components polarisations before
%ISP subtraction.
%

At this point, it must be stressed that the absence of correlation
between the interstellar and the individual polarisation {\em levels}
(see Figure~\ref{fig:pabvspmis}) could also result from an erroneous
estimation of the interstellar polarisation.  For instance, if the
probe stars are placed at a larger distance than the binary, i.e.,
behind the cloud, then the interstellar polarisation level will likely
be overestimated.  This point will be addressed in details in
sect.~\ref{subsec:calbin}.  However, even if the interstellar
polarisation level is overestimated, its orientation is likely to
remain correct.  Hence the absence of a complete correlation between
the various polarisation orientations (on the primary, the secondary
and the ISP) indicates that at least about 30\% of our measurements are
most probably not strongly influenced by the ISP.

\subsection{Polarisation measurements corrected for ISP}

For the remaining 18 sources, we use the local ISP
$\overline{Q}$ and $\overline{U}$ components and we compute the
calibrated polarisation of the central binary components A/B:
\begin{eqnarray} \label{eq:sub-uq} Q'_{A/B} & = & Q_{A/B} -
\overline{Q}\\ U'_{A/B} & = & U_{A/B} - \overline{U} \\ P'& =&
\sqrt{Q'^{2}+U'^{2}}\\ \Theta' & = & 1/2\arctan(U'/Q') \end{eqnarray}
Table~\ref{tab:pol-calib} lists the values of the ISP-corrected level
and PA for these sources.
\begin{table}[htb]
\caption[]{\label{tab:pol-calib}  ISP-corrected polarisations for all the remaining 18 sources where both the object and surrounding interstellar polarisation can be correctly evaluated }
\begin{flushleft}
\begin{tabular}{lllll}
\hline\hline
Source & $P'_A$ (\%) & $\Theta'_A (^o)$ & $P'_B$ (\%) & $\Theta'_B
(^o)$ \\
ESO H$\alpha\,29$ &  3.33 (0.12) & 27 (1)  &   1.75 (0.14)  & 171 (2)  \\
%
%Hen 3600        &  0.27 (0.04) & 9 (5)    &  0.02 (0.05)   & 171 (80)  \\
%
Sz 30              & 3.98 (0.14)    & 64 (1) & 3.88 (0.14)     & 64 (1)   \\
Sz2                 &  3.90 (0.13) & 22 (1) & 1.48 (0.07)   & 11 (1)  \\
Glass-I            &  4.67 (0.16) & 12.7 (0.5) & 12.3  (0.4)  & 26 (0.5)     \\
Sz 15              &  2.84 (0.13) & 107 (1) & 0.88  (0.16) & 35 (5)    \\
%
%Sz 48              & 3.62 (0.03) & 116 (0.3) & 3.70 (0.04) & 116 (0.3)  \\
%
Sz 62             &  0.12 (0.12) & 23 (28)  & 0.17 (0.12) & 43 (20)  \\
Sz 60             &  2.07  (0.08) & 165 (1) & 0.82  (0.05) & 169 (1.5)   \\
HO Lup          &  0.48 (0.08) & 20 (4)    & 0.89 (0.07) & 30 (2)   \\
Sz 116            &   3.5 (0.13) & 75 (1) & 3.73 (0.16) & 75 (1)   \\
SZ 81             &  0.7 (0.05)   & 82 (3)     & 0.80 (0.07) & 53 (2)     \\
 SZ 65            &  1.56 (0.14) & 65 (3) & 2.37 (0.3) & 39 (3)  \\
%
%WSB 20         &  2.84 (0.05) & 113 (1) & 2.85 (0.05) & 113 (1)   \\
%
WSB 18         &  1.77 (0.07)  & 90 (1)  & 1.76 (0.07) & 89 (1)   \\
WSB 26         &  3.52 (0.12) & 76 (0.5) &3.5 (0.12) & 76 (0.6)  \\
%
%WSB 19         &  1.09 (0.06) & 139 (2) & 1.26  (0.1) &152 (2)     \\
%
WSB 35         & 2.9 (0.1) & 132 (0.5) & 2.6  (0.1) & 134 (1)       \\
WSB 4           &   1.38 (0.06) & 102 (1) & 1.2   (0.05) & 108 (1)    \\
%
%WSB 42        &   3.10 (0.04) & 91 (1) & 11.4    (0.1) & 98 (1)    \\
%
WSB 28        &  1.28 (0.06) & 159 (1) &2.7   (0.3) & 164 (3)   \\
HBC 679      &  1.84 (0.07) & 9 (0.7) & 2.3 (0.24) & 10 (2)   \\
AS 353        &  1.88 (0.12) & 110 (2) & 1.54  (0.14) & 113 (2.5) \\
%

%%AS 205 A & 1.29 & 1.45& 0.20& 88.9 &2.9 \\ B & & 5.53& 1.06& 51.3
%%&4.6 \\ VV CrA A & 1.93 & 4.18& 0.25& 108 &1.7 \\ B & & 6.78& 1.73&
%%104.2 &7.3 \\
\hline
\end{tabular}
\end{flushleft}
\end{table}

\section{Discussion}
\label{sec:discussion}
   
\subsection{Spatial dependence of the instrumental polarisation}
\label{sec:poldep}

Patat \& Romaniello (2005) recently showed that FORS1 suffers from
variable instrumental polarisation across the field of view, following
a centrally symmetric pattern. A fit to the data shows the
polarisation level to vary radially as $0.057\,r^2$ (in \% with $r$ in
arcmin), from 0\% at the center up to $\approx 1$\% at
the corners. Such an instrumental pattern is of great concern in our
measurements as we use the surrounding field stars to estimate the
average value of the interstellar polarisation at the center of the
field where the binary object is located.

The instrumental polarisation level remains below 0.1\% within one
arcmin from the geometrical center of the detector.  In order to
estimate the effect of a spatial variation of the instrumental
polarisation in our data, we used three of our images where the
polarisation appears smooth (i.e., SZ2, WSB\,4, WSB\,20, see
figs.~\ref{fig:cartepol1} and \ref{fig:cartepol2}).  Assuming that the
actual interstellar polarisation is uniform across the field, we
computed the average polarisation of the objects in a circle of radius
1 arcmin from the center (excluding the central binary), and we
subtracted it from all the other measurements in the field in an
attempt to remove the (large) interstellar polarisation and isolate an
instrumental component.
\begin{figure}[htb]
    \begin{center}
\includegraphics[width=8 truecm]{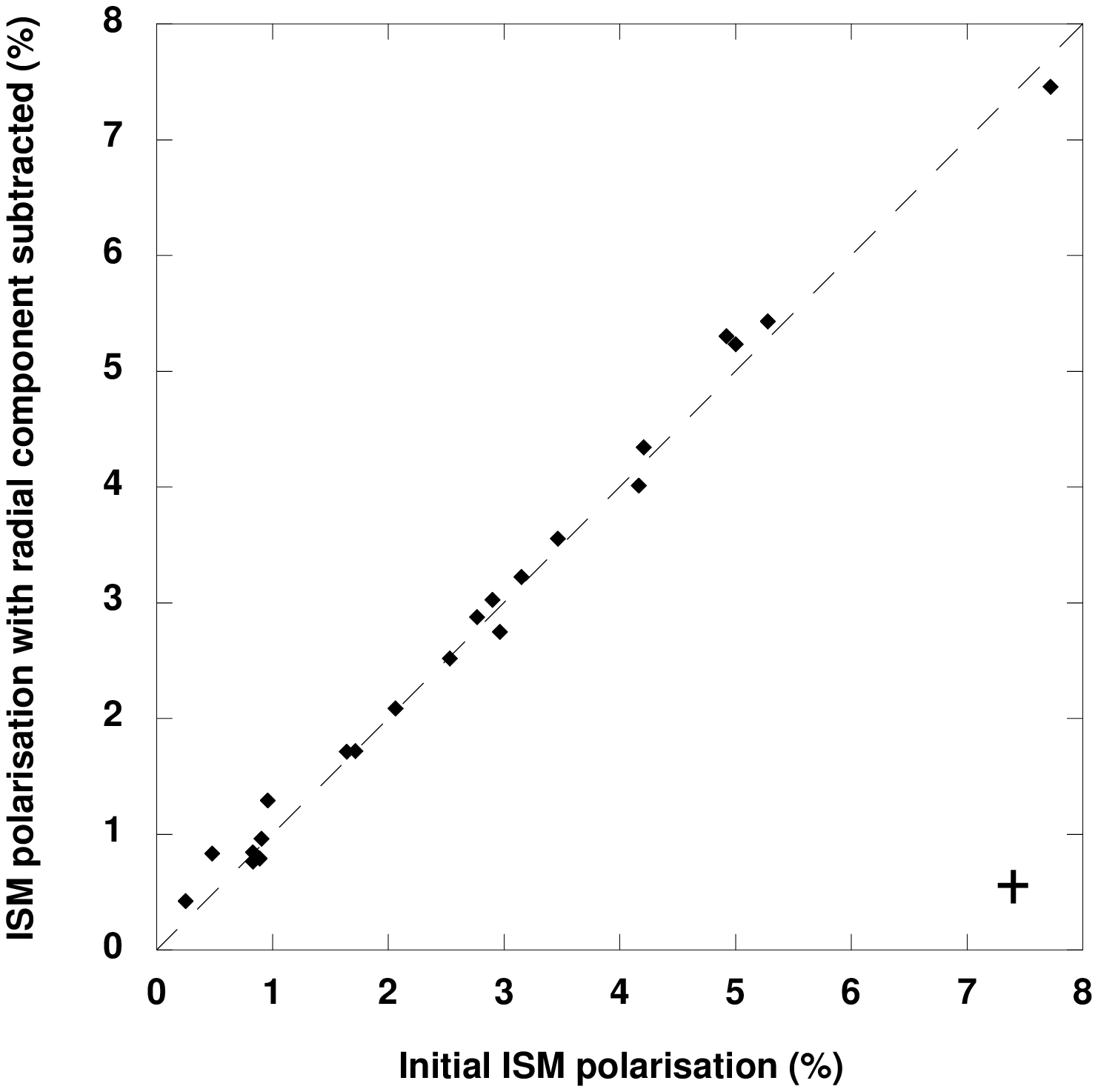}
\caption[]{\label{fig:comp-radial}  Comparison of the estimated 
interstellar polarisation with and without subtracting an instrumental
radial component. The cross at the lower right indicates the typical
error. The impact of intrumental polarisation on the measurements 
appears negligible.}
\end{center}
\end{figure}
Our results are consistent with an increase of instrumental
polarisation with distance from the center, increasing as $0.06\,r^2$.
We have then modeled a centrally symmetric instrumental polarisation
component with such a variation and verified its influence on the
parameters we extract from our images.  None of them is significantly
modified. As an example, Figure~\ref{fig:comp-radial} shows that this
effect on the estimation of the interstellar polarisation is very
small and can be neglected. This is so because the interstellar
polarisation in the clouds we observed is significantly larger than
the instrumental polarisation and because this instrumental
contamination is centrally symmetric, i.e., it cancels upon averaging
when numerous, well distributed stars accross the fields are used. We
are therefore confident that this instrumental effect does not modify
significantly our conclusions on the alignment of discs in young
binaries.

\subsection{Why would the secondaries be more polarised?}
   \label{subsec:secpol}

Figure~\ref{fig:pbvspa} shows that the measured polarisation level in
the secondary component is almost always equal or larger than that of the
primary, independently of the primary's polarisation level.
   
%Multiple scattering polarisation models show that when the
%polarisation is dominated by forward multiple scattering in the disc,
%the result is a larger polarisation level for a larger (more massive)
%disc.  If so, and if the disc mass is correlated to the central star
%mass, it is possible that in some sources, the primary is not the more
%massive object in the binary, and the apparent secondary likely
%possesses a larger disc, hence displays a larger polarisation than the
%apparent primary. However, even if some of our objects are
%misclassified as primary vs. secondary, it should not be the rule.

%
\begin{figure}[htbp]
    \begin{center}
\includegraphics[width=8 true cm]{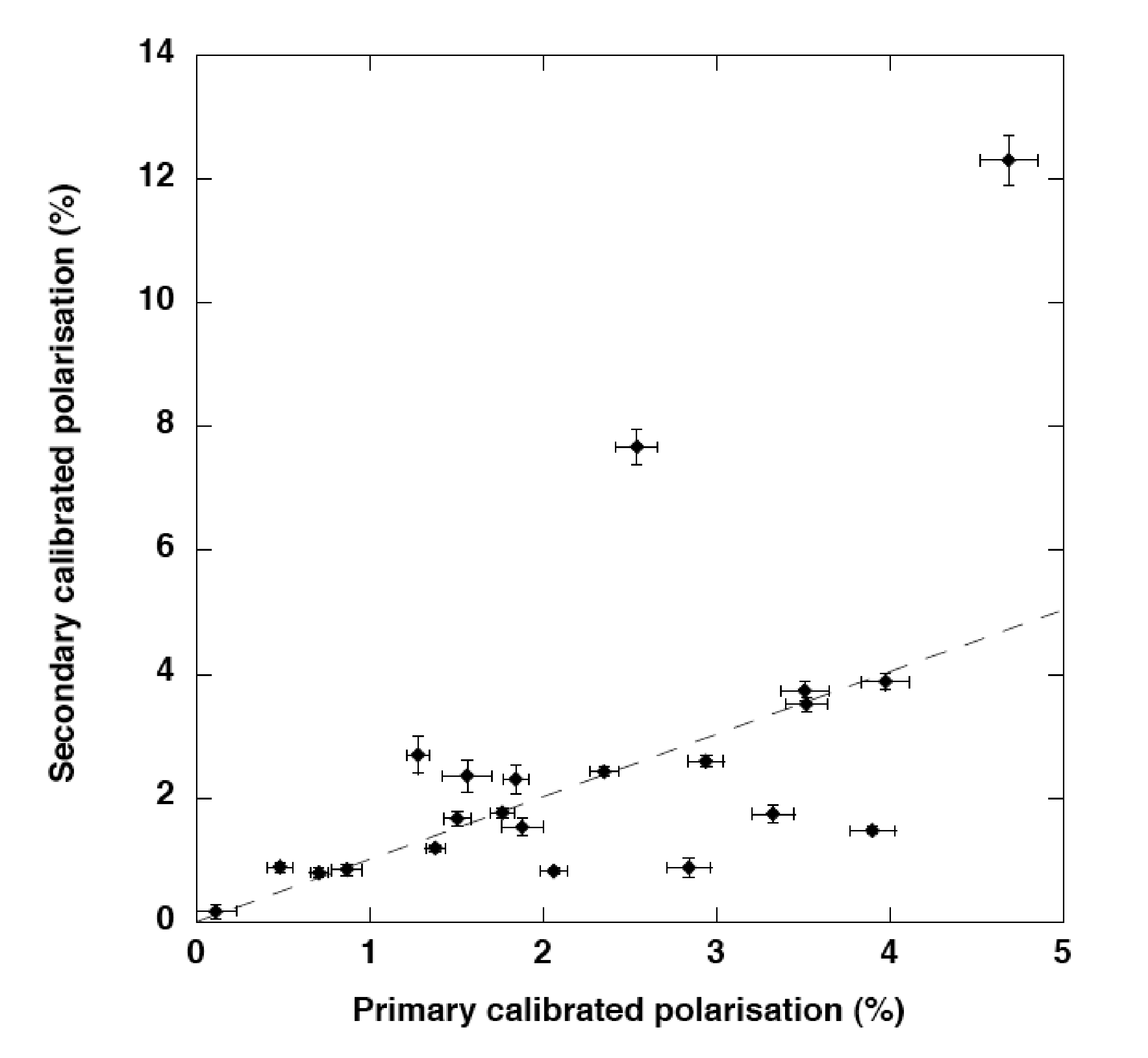}
\caption[]{\label{fig:piavspib}  ISP-corrected polarisation level in the secondary 
versus in the primary.}
\end{center}
\end{figure}

To verify that trend, Figure~\ref{fig:piavspib} contains a plot of the
{\em ISP-corrected } polarisation on the secondary versus the
primary. The error bars are larger due to the subtraction of the
ISP. After ISP subtraction the plot still suggests that a majority
of systems have similar {\em intrinsic} polarisations.
However, about a third of the systems show a significant
difference between the polarisation level in both components, but
contrary to Figure~\ref{fig:pbvspa}, there is no more tendency for one
of the component (B) to be statistically more polarised than the other (A).

A possible explanation of this effect is a difference in the relative
disc inclinations of the two components. In that case, a large
polarisation difference may occur as the more extinct component (by the
disc on the line-of-sight) will also be the more polarised (Monin et
al. 1998).

This situation is similar to the case of HK~Tau, where Duch\^ene et
al. (2003) have found that each component of the binary has a disc,
based on thermal emission at millimeter wavelengths, but they are not
parallel to each other as seen on optical images: the fact that the
(almost edge-on) secondary disc is visible when the primary one is not
can be explained by a relative inclination $\Delta i$ larger than
$15^o$. Such a difference is fully consistent with our results as
about one third of our sources show (ISP-corrected) PA differences
larger than $10^o$ (see fig.~\ref{fig:pbivspai}).

%However, one might doubt this cosmic conspiracy where the 
%classification of the binary 
%components is solely due to the inclination of their disc. 
%If indeed the mass difference is 
%not the dominant parameter, the fact that we measure 
%higher polarisations in the 
%secondaries may argue for  a variation of the physical characteristics 
%of the grains in 
%the disc. polarisation difference could come from  
%grain coagulation and / or 
%stratification in the disc. If grain growth is  enhanced in 
%larger discs due to the larger 
%pressure, the presence of larger grains will reduce the polarisation. 
%Similarly, if the 
%grains have began to settle on the mid-plane of larger discs, 
%the absence of grains in 
%the outer part of the disc will result in a lower polarisation. 
%The simultaneous study of 
%the polarisation level in young binaries hence brings a valuable 
%tool to tackle the 
%question of the primordial stages of planets formation. 
%A complete study of this effect is out of the scope 
%of this work, and will be published in a 
%forthcoming paper %(Pinte et al. 2005).

\subsection{Can we confidently remove the ISP component?}
\label{subsec:calbin}

In this section, we dicuss only the 18 binaries where both the central binary and the
surrounding ISP can be measured reliably.  However, even on very
regular fields like around SZ~116 or SZ~2, the actual interstellar
polarisation orientation might be correctly estimated, but its
amplitude can be overestimated if the probe stars are situated far
behind the cloud where the binary is placed.  

If we call $Q,U$ the true interstellar polarisation Stokes
parameters actually superimposed on the true
$Q^{o}_{A,B},U^{o}_{A,B}$ binary polarisation parameters (A: primary,
B: secondary), the estimators $\overline{Q},\overline{U}$ we compute
from the background star distribution can overestimate $Q,U$ so that :
$$ Q^{o}_{A,B} = Q_{A,B} - \alpha\, \overline{Q}$$
$$ U^{o}_{A,B} = U_{A,B} - \alpha\, \overline{U}$$
Where $0<\alpha<1$ if we assume that the background stars can be 
anywhere but between the source and the observer. 

Taking $\alpha=0$ is equivalent to ignore the influence of the local
interstellar polarisation, hoping that it will not statistically
change the result of the analysis.  This is the choice made by
previous authors who did not measure the IS polarisation.  Using
$\alpha = 1$ is the choice made so far in this paper.  

We now examine the intermediate case: $0<\alpha<1$. This is necessary
whenever the ISP value is overestimated, e.g., by measuring stars
much farther in the background.

To check the influence of an overestimation, one can look for a unique
value of $\alpha$ allowing to build {\em simultaneously} $Q_{A},U_{A}$
and $Q_{B},U_{B}$ from $\overline{Q},\overline{U}$.  The best
estimation of $\alpha$ that would simultaneously null $Q_{X},U_{X}$
is: $$ \alpha (X) = {Q_{X}\overline{Q} +
{U_{X}\overline{U}}\over{{\overline{Q}}^{2}+{\overline{U}}^{2}}}$$
($X\equiv A,B$).

The goal is to find a value of $\alpha$ that would null both
polarisations. In that case, the measured polarisation of the two
components of the binary is very likely to be entirely of interstellar
origin. Three binaries are found in the sample where the polarisation of 
both components can be simultaneously cancelled by the same fraction of interstellar polarisation.  Those binaries are: Sz~30
($\alpha (A) = 0.25$ ; $\alpha(B) = 0.27$); Sz~62 ($\alpha(A) = 0.96 ;
\alpha(B) = 0.95$); Sz~116 ($\alpha(A) = 0.02 ;
\alpha(B) = 0.05$).  In the latter case, the initial polarisation of
the binary is very weak and if it is not entirely from interstellar
origin, it can hardly be used to study disc alignment anyway.  In
these 3 cases, we can not disentangle the polarisation of the sources
from a possible interstellar origin. They are removed for any further
statistical analysis. For the other systems, no common value of
$\alpha$ can be found, suggesting differences in the intrinsic
polarisation of each component.

\begin{figure}[htb]
   \begin{center}
\includegraphics[width=8 truecm]{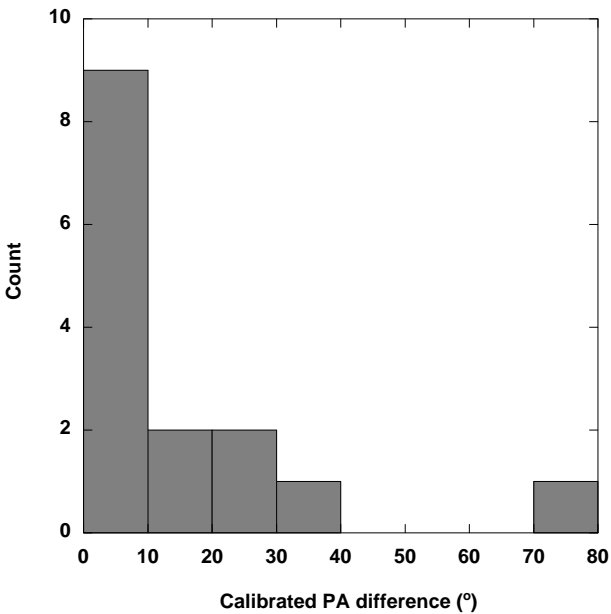}
\caption[]{\label{fig:pbivspai} Histogram of the calibrated position 
angle difference in the 15 remaining binaries. }
\end{center}
\end{figure}

For the discussion below we choose to keep $\alpha=1$ for the
remaining 15 sources. In Figure~\ref{fig:pbivspai}, we have plotted
the histogram of the ISP-corrected position angle difference between
the primary and the secondary.  60\% (9/15) of the sources show
alignment better than 10$^{o}$, and 73\% (11/15) better than 20$^{o}$.
Only 4 sources show angle difference exceeding 25$^{o}$, and there is
a unique case (Sz~15) where the polarisations are almost perpendicular
to each other.

%It is worth noting that all the sources with a calibrated 
%PA difference larger than 25$^o$ have a 
%secondary more polarised than the primary.

The error bars are of the order of a few degrees on the position angle
($\le 5^{o}$). The good correlation between both polarisation position
angles suggest that if the polarisation PA correctly traces the
intrinsic disc orientation, then there is a strong alignment tendency
between discs in binaries during the T Tauri phase of early stellar
evolution.

\subsection{Alignment versus separation}
 
\begin{figure}[htb]
    \begin{center}
\includegraphics[width=8 truecm]{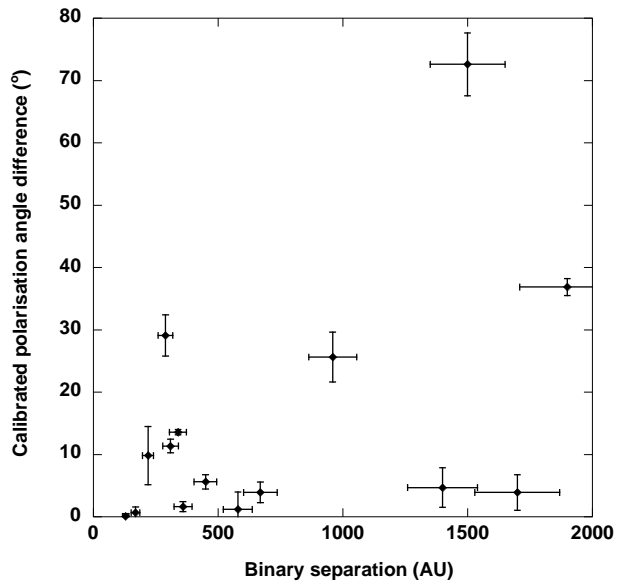}
\caption[]{\label{fig:disc-pa-sep}Discs orientation difference versus
separation in AU. The error bars on the position angles are determined
in our data reduction pipeline; we plot a 10\% error on the
distance determination.}
\end{center}
\end{figure}

In this section we compare the polarisation position angle difference 
with the projected linear binary separation. The result is presented 
in Figure~\ref{fig:disc-pa-sep} for the
remaining 15 sources of the sample. No clear correlation appears but
the distribution of angle differences is consistent with a larger
difference for wider sources, although this could also reflect random
alignment for non physical pairs.  If the general trend is real, this
is consistent with the results of Bate et al.(2000) that show
that due to the shortness of the alignment time-scale, strongly
misaligned discs are only likely to occur in binaries with separations
larger than 100~AU.  In our sample, we find that in 80\% of the
binaries with separation less than 700\,AU, the discs are aligned to
better than $15^o$.  In this number, we have counted out Sz~2 because a
strong polarisation level difference exists between the primary and
the secondary, that can be interpreted in terms of inclination angle
difference.

\subsection{Alignment pattern and timescale in various SFR}
\label{subsec:var-sfr}
 
Our results are in line with previous estimates for wide binaries
(Monin et al.  1998) and for binaries with similar separations in
Taurus (Jensen et al.  2004).  Thus, it appears that a disc alignment
tendency is a common phenomenon in young SFR. Bate et al. (2000)
suggest that in very dense star forming environments, misaligned discs
could occur due to close dynamical interactions with other cluster
members.  The lack of misaligned discs in our results suggest that
either the SFR's we surveyed were not dense enough for strong
misalignment to persist, or the disc alignment timescale is short
enough that whatever the low age of the SFR, the discs
always had time to re-align.

We also find that if closely aligned discs exist, there are also pairs
that are misaligned, with PA differences larger than 30$^{o}$. This is
also consistent with results from Bate et al. (2000), namely that the
alignment timescale depends on the degree of initial misalignment:
strongly misaligned discs take very little time to reduce their large
difference in orientation but take a much longer time to reach perfect
alignment, hence the existence of a tail in the distribution of
position angle difference.

%\subsection{Disc alignment in Southern SFR vs Taurus}
%
%M\'enard \& Duch\^ene (2004) have re-examined the question of the
%alignment of discs with the local magnetic field in the Taurus cloud.
%They found that the overall CTTS population is oriented randomly with
%respect to the magnetic field.  While differences may exist between
%different SFR due to various interplay between the cloud magnetic
%field and other physical properties of the region, our sample has been
%selected to include CTTS and emission line stars with active accretion
%discs, hence probable outflow activity also, i.e, active jets. Thus
%our finding of a common binary disc orientation perpendicular to the
%local ISP, tracing the magnetic field, appears consistent with
%M\'enard \& Duch\^ene (2004) study.

\section{Conclusion}
\label{sec:conclusion}

We have presented the results of a polarimetric survey on 23 southern
pre-main sequence binaries closer than 2000~AU in separation, half of
them being closer than 340~AU.

We have obtained $6.8'\times 6.8'$ polarimetric maps around all the
binaries.  The observations allow to estimate the linear polarisation
of the central binary and an estimation of the local interstellar
polarisation from surrounding background stars. We find that
estimating the interstellar polarisation on the central binary is a 
difficult task. In particular, `by eye' estimation of the local
polarisation from the literature leads to significant errors.
% as the only
% physical quantities that can be evaluated and subtracted from the
% central object measurements are the Stokes parameters $Q$ and $U$.

In every binary, we have estimated an ISP-corrected polarisation for
each component. We find that the polarisation levels for the primary
and the secondary are correlated, although in some cases a strong
difference exists between the two.  Such a difference can be
interpreted in term of a difference in the inclination of the discs
relative to the line-of-sight.  Indeed, the orientation we determine
are projection on the plane of the sky, so for a given binary, we
cannot rule out that discs have different inclinations, even if the
position angle of the two components are similar. However, the fact
that the calibrated polarisation levels remain highly correlated
between primaries and secondaries is consistent with both discs
sharing also similar orientations toward the line of sight.  This is
in agreement with Wolf et al. (2001) who calculated that a position
angle difference distribution peaking toward zero is consistent with a
disc alignment tendency.

We find that after interstellar polarisation subtraction, 73\% (11/15)
of the disc polarisation, hence the putative rotation axes of the
components, are aligned within $20^{o}$, a result consistent with
previous work (see Monin et al., 1998; Donar et al.  1999; Jensen et
al.  2000; Wolf et al.  2001; Jensen et al.  2004).  This proportion
falls to 60\% (9/15) when one takes into account the large
polarisation level difference that exists between components with
aligned discs in the plane of the sky. This result can be interpreted
in terms of large difference between the inclination angles of the
discs. In any case, this large proportion of aligned discs may reflect a
primordial alignment during star formation.

%This primordial alignment hypothesis is supported by our
%finding that the primary intrinsic polarisation PA is preferentially
%oriented at $90^{o}$ from the interstellar polarisation PA. If the
%interstellar polarisation mechanism is dominated by preferential
%absorption of background starlight by elongated grains aligned on the
%local magnetic field, and if disc integrated polarisation is aligned
%with the disc equatorial plane, this result supports the scenario
%where discs in binaries form by collapse oriented preferentially
%parallel to the magnetic field lines.

\begin{acknowledgements}

We thank the ESO support staff for their help during the observations.
We also thank Gaspard Duch\^ene and Christophe Pinte for enlightening
discussions, and an anonymous referee for useful comments that helped
improve the quality of the paper. We warmly thank F.~Patat \& M.~Romanielo for a copy of their paper prior to publication that helped us estimate the influence of the instrumental polarisation on our measurements.
We acknowledge the {\sl Programme
National de Physique Stellaire} (PNPS) of CNRS/INSU for supporting the
star formation programme at LAOG. This research has made use of NASA's
Astrophysics Data System Bibliographic Services, and the SIMBAD
database, operated at CDS, Strasbourg, France.

\end{acknowledgements}

%-BIBLIOGRAPHY------------------------------------------------------------------
%\bibliographystyle{apj}

%%%%% -----------------------------------------

\end{document}